\documentclass[aps,pra,reprint,10pt,a4paper]{revtex4-1}
\usepackage[utf8]{inputenc}
\usepackage[T1]{fontenc}
\usepackage{amsfonts}
\usepackage{amssymb}
\usepackage{amsmath}
\usepackage{amsthm}
\usepackage{graphics}
\usepackage{graphicx}
\usepackage{braket}

\begin{document}


\title{Spreading Nonlocality in Quantum Network}
\author{Ratul Banerjee, Srijon Ghosh, Shiladitya Mal, Aditi Sen (De)}
 \affiliation{Quantum Information and Computation Group,
Harish-Chandra Research Institute, HBNI, Chhatnag Road, Jhunsi, 
Allahabad 211 019, India}

\begin{abstract}
Starting from several copies of bipartite noisy entangled states, we design a global and optimal local measurement-based protocol in one- and two-dimensional lattices by which any two or more prefix sites can be connected via entanglement. Production  of bipartite as well as  multipartite entangled states in a network is verified in a device independent way through the violation of Bell inequalities with two settings per site and with continuous range of settings. We also note that if the parties refuse to perform local measurements, the entanglement distribution scheme fails. We obtain critical values of noise allowed in the initial state so that the resulting output state show nonlocal correlation in different networks with arbitrary number of connections. We report that by employing our method,  it is  possible to create a Bell-violating multipartite entangled state from non-Bell violating bipartite states in an one-dimensional lattice with minimal coordination number being six. Such a feature of superadditivity in violation can also be observed in a triangular two dimensional lattice but not in a square lattice.

\end{abstract}

\maketitle

\section{Introduction}
\label{sec_intro}
At the age of internet,  ability to  share information among  arbitrary number of  parties situated in different locations is  the basic building block for a communication network \cite{cirac'97, duan'01, winter'09, kimble'08}. The information distributed can both be classical as well as quantum in nature. In both the cases, it was shown that entangled states shared between the senders and the receivers can enhance the performance of the protocol, which cannot be achieved by unentangled states \cite{HHHHRMP} .  Therefore, the first step towards  establishing communication network is to generate high entanglement in bipartite as well as multipartite states connecting different sites. To achieve this goal, several protocls have been developed which include quantum repeaters \cite{briegel'98, dur'99} based on entanglement distillaton from noisy shared states followed by entanglement swapping \cite{zzhe'93, Bose98}, entanglement percolation \cite{acin'07, perse'08, cuquet'09, broadfoot'09} originated from the notion borrowed from statistical mechanics,  k-pair communication problem \cite{winter'09}. 

%

A network can be defined in an one- and a two-dimensional lattices with different geometrical structure -- the edges of the lattice is covered by bipartite states and depending on the lattice, several edges are connected through each vertex or node, determining the coordination number of the lattice (see e.g. \cite{acin'07, jiang'08} and references therein). Initial states covering the lattice can also be a ground or a canonical equlibrium  or an evolved state at certain time of a suitable Hamiltonian. A prominent example is the resonating valence bond states in which every nodes are connected by a singlet \cite{RVB}.
Typically, suitable joint measurements performed at each node can create a multipartite quantum correlated state between prefix sites \cite{acin'07} which can be used later for  quantum information processing tasks. Note that such a notion has also been used to build measurement-based quantum computer  \cite{onewayQC}. Performance of all these tasks are measured, for example, by localisable entanglement \cite{popp'05}, singlet conversion probability \cite{acin'07}, gate fidelity \cite{onewayQC}. 


One of the most counter-intuitive features of entangled state is that it exhibits a kind of 'nonlocal' effect. Specifically, it means, no local-realistic model can account for all the correlation emerging from local measurements on entangled states \cite{Bell'64, chsh'69, Bellreview}. Although not all entangled states violate Bell inequality \cite{werner'89}, Bell test \cite{belltest1, belltest2} turns out to be the device independent certification of entangled states. Violation of Bell inequalities are shown to be useful in quantum cryptography \cite{crypto}, random number generation \cite{rand} etc.  Moreover, several preprocessing protocols were prescribed to probe violation of Bell inequality for states which do not respond to Bell test. In  this direction, it was shown in the seminal paper by Sandu Popescu  \cite{popescu'95} that  local filtering can help to reveal the nonlocality known as hidden nonlocality (see also \cite{gisin'96, hirsch'13}). Other activation protocols involving  Bell test with multiple copies  were proposed \cite{peres'96, masanes'06}. In a similar spirit, violation of local realism was demonstrated in a multisite domain employing entanglement swapping --- initial seven or more copies of non-Bell violating Werner states, forming a star network, leads to the final multipartite state which violates functional Bell inequality \cite{zukowski'93}. It is known as superadditivity in nonlocality \cite{aditi'05} (cf. \cite{cavalcanti'11}).

In this work, we propose a framework of a quantum network based on global and local measurement, in which bipartite and multipartite  entangled states  are  generated in two or more arbitrary  prefixed sites in one- and two-dimensional lattices.  Specifically, the entire lattice is covered by the Werner states and  depending on the sites which we want to connect,  minimum number of joint and local measurements  are implemented resulting in entangled state which can be certified through Bell test \cite{Bell'64, chsh'69, mermin'90, bk'93, zukowski'93}. For bipartite case, we consider  Clauser-Horne-Shimony-Halt (CHSH) inequality \cite{chsh'69} while for multipartite states, we employ Mermin-Belinskii-Klyshko (MBK)\cite{mermin'90, bk'93} and functional Bell (FB) inequalities \cite{zukowski'93}.

 In an one-dimensional (1D) lattice, we report that for exhibiting superadditivity, the minimum coordination number required is six. For a fixed number of nodes, the value  of  critical noise allowed in the initial state, leading to nonlocal multiparty state, is determined  for different coordination number.  Similar analysis has also been performed  with varying number of nodes for a fixed coordination number.  We also find the maximum amount of noise accepted in the initial state resulting in nonlocal correlations in the output state in square and triangular two-dimensional (2D) lattices.  Moreover, we observe that the superadditivity in violation can only be shown in a two-dimensional lattice having lowest co-ordination number six, eg. in a triangular lattice but not in a square lattice. Specifically, we show that for a fixed number of joint and local measurements, it is always possible to find minimal number of nodes for which superadditivity in nonlocality can be exhibited. 

We organise the paper in the following way. In Sec. \ref{sec_neccessary}, we first introduce the notion of violation of Bell inequality in a multipartite domain when some of the parties collaborate, which we call as localizable nonlocality. We then consider  localizable nonlocality in a star network in Sec. \ref{sec_block} and find critical noise required to obtain violation in the output state. Sec. \ref{sec_network} is devoted to  the results obtained for 1D and 2D  lattices and we conclude with discussion in Sec. \ref{sec_conclu}.

\section{Violation of Bell inequality with collaboration}
\label{sec_neccessary}

Let us consider that $N$ parties share a multipartite state, $\rho_{N}$, and among $N$ parties, $m$ number of parties collaborate by performing local projective measurements $\{M_i\}$ in their respective part of the state, which leads to an ensemble $\{p_i,\rho^{i}_{N-m}\}$. Here $p_i$ is the probability pertaining to a specific outcome combination obtained by $m$ parties who want to collaborate with other $N-m$ parties to perform Bell test. We define average value of Bell expression of the post-measurement ensemble consisting of $N-m$ party state, $\{p_i,\rho^{i}_{N-m}\}$ as '{\it localisable nonlocality}' (LNL).
\begin{eqnarray}
\mathcal{L}_{NL} = \max_{\{M_i\}}\sum_{i} p_i \mathcal{BV}(\rho^{i}_{N-m}),
\end{eqnarray}
where $\mathcal{BV}$ indicates the amount of violation of appropriate Bell inequality and maximization is performed over the set of all local measurements, $\{M_i\}$ by $m$ number of parties. Eg., when $N-2$ parties  measure locally on their respective subsystems, the violation of CHSH inequality of the resulting bipartite state is studied, thereby certifying the entanglement of the bipartite state in a device-independent manner.  On the other hand, when $N-m > 2$,  the output is a multipartite state and we analyse the violation of MBK \cite{mermin'90, bk'93} and FB  \cite{zukowski'93} inequalities. Before investigating $\mathcal{L}_{NL}$ in network, let us briefly discuss the Bell operators that we will use in this paper.

{\bf{Condition for violation of CHSH inequality:}} As stated earlier, let us first describe the CHSH inequality and its violation for two spin-half particles \cite{chsh'69}. Suppose a bipartite state $\rho_{AB}$ is shared between two spatially separated observers, say, Alice and Bob. They both can choose to perform a dichotomic measurement at a time from different set of two observables. CHSH inequality puts a restriction on a particular algebraic expression imposing locality and reality assumptions. It involves correlation between local measurement statistics of Alice and Bob, i.e.,
\begin{eqnarray}
\mathcal{B}\equiv |\langle A_1 B_1\rangle +\langle A_1 B_2\rangle +\langle A_2 B_1\rangle -\langle A_2 B_2\rangle| \leq 2.
\end{eqnarray}
Here $\langle A_i B_j\rangle =\mbox{Tr}(A_i\otimes B_j \rho_{AB})$ is the correlation between measurement outcomes $a_i$ and $b_j$ for the measurement $A_i$ and $B_j$ performed by Alice and Bob respectively.
For an arbitrary two-qubit state, the maximal violation of the CHSH inequality in terms of state parameters were derived \cite{horodecki'95}. In particular, maximal violation of local realism in this case can be written in terms of  correlation matrix, $T$, whose elements are defined as
\begin{eqnarray}
T_{ij} = \mbox{Tr}[\rho_{AB}\sigma_{i}\otimes\sigma_{j}]
\end{eqnarray}
where $\sigma_{i}\,\, (i=x, y, z)$ are Pauli matrices. A state is considered to violate Bell inequality if 
\begin{eqnarray}
\mathcal{M}(\rho_{AB}) > 1
\end{eqnarray} 
where $\mathcal{M}$ is the sum of the two maximum eigenvalues of $T^{\dagger}T$. Hence, maximal Bell violation is quantified as
\begin{eqnarray}
BV(\rho_{AB})= M(\rho_{AB}) -1,
\end{eqnarray}
and finally $\mathcal{L}_{NL}$ reads as
\begin{eqnarray}
\mathcal{L}_{NL}^{CHSH} = \max_{\{M_i\}}\sum_{i=1}^{2^{N-2}} p_i BV(\rho^{i}_{AB}),
\end{eqnarray}
where $\rho_{AB}^i$ is obtained after performing local measurements on $N-2$ parties of an initial state $\rho_{N}$. 

{\bf {Violation of Mermin-Belinskii-Klyshko inequality:}} It is a multiparty correlation-function Bell inequality in which each party can choose to measure from a set of two observables \cite{bk'93}. In a $N$-partite state,   when $m$ number of parties perform local measurements, the violation of MBK inequality for the rest of the ($N-m$)-qubit  state is given by the expectation value of the MBK operator \cite{scarani'01, dur'01},
\begin{eqnarray}\label{mki}
    B_k & = & \frac{1}{2} B_{k-1}\otimes (\sigma_{a_{k}}+\sigma_{a'_{k}}) \nonumber \\
    & + &  \frac{1}{2} B'_{k-1}\otimes (\sigma_{a_{k}}-\sigma_{a'_{k}}),
\end{eqnarray} 
where $a_k$ and $a'_k$s are the two vectors on the unit sphere, that indicate possible measurement directions of the corresponding party. $B_k$ is obtained recursively from $B_{k-1}$ and $B'_k$ is obtained from $B_k$ by interchanging all the $a_k$s by $a'_k$s.
 A state $\rho_{N}$, is said to violate MBK inequality if the average value of this operator becomes greater than $1$ i.e.,
  \begin{eqnarray} 
  |\mbox{Tr}[B_{N}\rho_{N}]|>1,
  \end{eqnarray}
  and the  corresponding LNL can be computed as 
  \begin{eqnarray}
\mathcal{L}_{NL}^{MBK} = \max_{\{M_i\}}\sum_{i=1}^{2^{m}} p_i  |\mbox{Tr}[B_{N-m}\rho_{N-m}]|_{i}  - 1.
\end{eqnarray}

{\bf{ Violation of functional Bell inequality:}} 
Like MBK inequality, this  multisite inequality  is based on Schwartz inequality.  Instead of two settings at each site like CHSH and MBK inequalities, it considers a set-up involving continuous range of settings \cite{zukowski'93}.
Let $ G_{n}$ be an local observable at the nth party $(n = 1, . . . , N)$, and each of them depends on some parameter $\eta_n$.
Based on these measurements, $G_n$s, we define correlation function as 
\begin{eqnarray}
C_{QM}(\eta_{1},....\eta_{N}) =Tr(\rho_{N}G_{1}...G_{N}),
\end{eqnarray}
and the corresponding correlation admitting local hidden variable model, with the distribution of local variable, denoted by $v(\lambda)$, reds as 
\begin{eqnarray}
C_{LHV}(\eta_{1},....\eta_{N})= \int d\lambda v(\lambda)\prod_{n=1}^{N}\mathbb{I}_{n}(\eta_{n},\lambda).
\end{eqnarray}
Here  $I_{n}(\eta_{n}, \lambda)$ is the predetermined measurement result of $G_n$ for $\lambda$. To show $C_{QM}\neq C_{LHV}$, one can use the basic principle of Schwartz inequality, and so we compute
\begin{eqnarray}
\langle C_{QM}|C_{LHV}\rangle = \int d\eta_{1}...d\eta_{N}C_{QM}(\eta_{1},....\eta_{N})C_{LHV}(\eta_{1},....\eta_{N}) \nonumber
\end{eqnarray}
and the
\begin{eqnarray}
\left \| C_{QM} \right \|^{2} = \int d\eta_{1}...d\eta_{N}(C_{QM}(\eta_{1},....\eta_{N}))^{2},
\end{eqnarray}
which finally leads to $C_{QM}\neq C_{LHV}$.

 In this case, to calculate localized nonlocality,  we first consider the average overlap of  $ C_{QM}$ and $ C_{LHV}$ over $2^m$ outcomes after m parties perform local measurements. Suppose we can show 
 \begin{eqnarray}
\sum_{i=1}^{2^m}p_{i}\langle C_{QM}|C_{LHV}\rangle_{i} \leq H,
\end{eqnarray}  
where H depends on local measurement parameters of m parties.  
On the other hand, for a given outcome $i$, we can get $\left \| C_{QM} \right \|^{2}_{i}$ calculated for a post-measurement state of $(N-m)$ parties. Finally, we have
\begin{eqnarray}
\mathcal{L}_{NL}^{FB} = \max_{\{M_i\}}\sum_{i=1}^{2^{m}}p_{i}\left \| C_{QM} \right \|^{2}_{i} \nonumber -H
\end{eqnarray} 
Here $\langle C_{QM}|C_{LHV}\rangle_{i}$ (overlap of $C_{QM}$ and $C_{LHV}$ for the $i$th outcome) and $\left \| C_{QM} \right \|^{2}_{i}$ (the norm of $C_{QM}$ also with measurement result $i$) are defined as 
 \begin{eqnarray}
\langle C_{QM}|C_{LHV}\rangle_{i} = \int d\eta_{1}...d\eta_{N-m}C_{QM}(\eta_{1},....\eta_{N-m})_{i}\nonumber\\
C_{LHV}(\eta_{1},....\eta_{N-m})_{i},
 \end{eqnarray}
 \begin{eqnarray}
 \left \| C_{QM} \right \|^{2}_{i} = \int d\eta_{1}...d\eta_{N-m} (C_{QM}(\eta_{1},....\eta_{N-m})_{i})^{2}.
\end{eqnarray}

\section{Localisable nonlocality in star network: A building block}
\label{sec_block}
Before detecting entanglement via Bell test in a lattice, let us first fix the operations that we are going to perform for establishing connection between any two or more number of nodes in a network. In this section, we study the violation of Bell inequality in a geometry which turned out to be a building block (unit) of an entire netwok. 

\begin{figure}
\includegraphics[height=6.0cm,width=8.0cm]{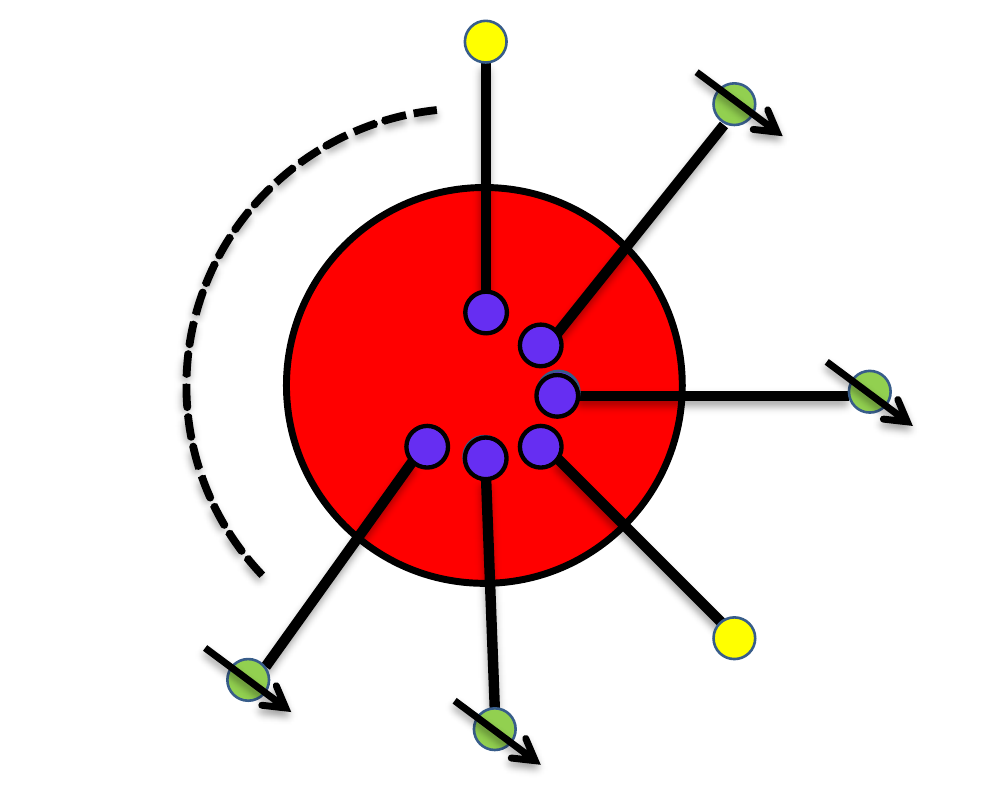} 
\caption{(Color online) Schematic diagram of a single block. It consists of N copies of $\rho_{W}$ states. An $N$-party GHZ-measurement is performed on the $N$ qubits situated in the centre marked as purple circles while local measurements are carried out on the (green) qubits, marked with arrow. Our aim is to produce an entangled state  between the yellow qubits.}
\label{block}
\end{figure}

Suppose, \(2N\) number of parties, $A_{i}$ and $B_{i}$ $(i = 1,2...N)$  sharing N identical copies of arbitrary bipartite state among them as shown in the Fig. \ref{block}. Let us consider a scenario in which all the $A_{i}$s are assumed to be situated in one place and hence we can replace them by an observer, say, Alice (A), while $B_{i}$s are located in distant positions forming a star network \cite{aditi'05}. Alice performs a projective joint measurement on the \(N\)-parties in her possession and consequently, a multipartite entangled state, $\rho_{N}$ is created among other distant \(N\) sites. 

We consider a scenario in which all the $A_i$ and \(B_i\) shares \(N\) identical copies of the Werner state, given by 
\begin{equation}
\rho_W = p |\psi^-\rangle \langle \psi^- | + (1-p) \frac{I}{4},
\end{equation}
where \(p\in (0,1)\). We know that it is entangled for \(p > 1/3\) while it violates CHSH inequality when \(p > 1/\sqrt{2}\) \cite{werner'89}.

As previously mentioned, $A$ measures in a $N$-qubit basis consisting of  Greenberger- Horne-Zeilinger state (GHZ) \cite{GHZ} in the center of the star. If one of the outcomes, say, $\frac{1}{\sqrt{2}}(|00..0\rangle + |11...1\rangle)$ occurs, the state shared between $B_{1}, \ldots, B_{N}$ is projected to an entangled $N$-qubit state.
E.g., if the initial state is $\rho^{\otimes3}_{W}$, $A$ performs measurement in the $\{\frac{1}{\sqrt{2}}(|000\rangle\pm|111\rangle), \frac{1}{\sqrt{2}}(|001\rangle\pm|110\rangle), \frac{1}{\sqrt{2}}(|010\rangle\pm|101\rangle),  \frac{1}{\sqrt{2}}(|100\rangle\pm|011\rangle)\}$-basis, which we call the GHZ-basis, easily extended to arbitrary number of qubits. It results in an output state of the form
\begin{eqnarray}\label{sform}
&& q_{1}|GHZ\rangle\langle GHZ| + q_{2} \frac{I}{8} \nonumber\\
&+&
 (1-q_{1}-q_{2})(|000\rangle\langle 000| + |111\rangle\langle 111|).
\end{eqnarray}
 Note that the probability of obtaining any of the outcome by the joint measurement of A is equal for the initial state, $\rho^{\otimes3}_{W}$.
In general, when the input state is  $\rho^{\otimes N}_{W}$, the form of the output state after the joint measurement by $A$, remains same  as in (Eq.\ref{sform}), which is known as $X$- type state having  nonvanishing diagonal and two cross diagonal terms. In particular, if one writes the final state in the computational basis, we find non-vanishing coefficients appearing in all the diagonal and in two off-diagonal terms, which are $|00...0\rangle\langle 11...1|$ and $|11...1\rangle\langle 00...0|$. Interestingly, we notice that in the evaluations of the Bell-CHSH, the MBK and the FB inequalities, the off-diagonal terms play a crucial role which we investigate carefully.

In order to establish nonclassical correlation in the network in terms of violation of CHSH, MBK or FB inequalities, we invoke two different strategies.
In both the scenarios, we assume, $N$ parties perform joint measurement and the output state is shared between N-parties which may face two situations.
Our aim is to produce a state, $\rho_{N-m}$, for which Bell violation is maximised over all local projective measurements. Note that we here consider only rank-1 measurement. In particular, we divide the entire protocol into the following three steps: 

{\bf Step 1:} We start with $N$ copies of a given bipartite states forming a star network. Alice performs a global measurement on $N$ parties in the center and consequently creating a $N$-partite entangled state.

{\bf Step 2:} Any $m$ number of parties either can perform local measurements or leave the protocol. It leads to the following cases:

\begin{enumerate}

\item {\bf Non-collaborative strategy:} $m$ number of parties leave the protocol without any measurements, i.e., they do not collaborate with $N-m$ parties. Mathematically, $\rho_{N-m}=Tr_{1 \ldots m}\rho_{N}$.
\item {\bf Collaborative strategy:} Among $N$ parties, $m$ parties collaborate  in a sense that they perform local projective measurements on their subsystems, so that an output  state of $N-m$-party is produced.
\end{enumerate}

{\bf Step 3:} We finally evaluate the violation of Bell inequality  of $N-m$-party state. 

In the second step, if strategy 1 is followed, we call the associated nonclassical correlation certified via violation of local realism as $N-m$ '{\it reduced nonlocality}' while if strategy 2 is followed, we call '{\it localisable non-locality}'. As we will see, the above scenario, especially the collaborative strategy can help  to spread entanglement over a large distance in a network from initial bipartite noisy entangled states.

{\bf No reduced nonlocality:}
If the initial state shared between $2N$ parties is the Werner state, we can easily find that non-collaborative strategy leads to a $(N-m)$-party state which is separable. Therefore, non-collaborative strategy is not suitable for spreading nonclassicality. In rest of the paper, we only concentrate on the collaborative strategy. 

\subsection{Locating bipartite nonlocality}
Suppose among m parties, $l$-th party performs a local measurement in the $\{|\pm\rangle_l\}$-basis, given by
\begin{eqnarray}
|\pm\rangle_l = \cos\theta_l|0\rangle \pm e^{-i\phi_l}\sin\theta_l|1\rangle,
\end{eqnarray}
where $l=1,2,...,m$. There are $2^m$ possible outcomes of the measurements. Corresponding to each outcome, we evaluate the maximum possible violation of Bell inequality for remaining $N-m$-partite state. The post selected state, $\rho_{N-m}$, takes the same form as the $N$-partite state with modified coefficients given in Eq. (\ref{sform}) provided the initial Werner states are projected by the joint measurement.
As mentioned earlier, in all the Bell expression considered here, the off diagonal terms of the density matrix, $\rho_{N-m}$, are important and are explicitly given by
\begin{eqnarray}
\langle 00\ldots 0|\rho_{N-m}| 11 \ldots 1\rangle_{i} = \pm\frac{p^{N}[e^{+ i\sum_{1}^{m}\phi_{l}}\prod_{1}^{m}\sin\theta_{l}]}{2(1- f_{i}(\theta_{1},\theta_{2},\ldots\theta_{m}))},\\
 \langle 11 \ldots 1|\rho_{N-m}| 00 \ldots 0\rangle_{i} = \pm\frac{p^{N}[e^{- i\sum_{1}^{m}\phi_{l}}\prod_{1}^{m}\sin\theta_{l}]}{2(1- f_{i}(\theta_{1},\theta_{2}, \ldots\theta_{m}))}.
 \end{eqnarray}
Probability of getting a particular outcome is a function of local measurement parameters, namely 
\begin{eqnarray}
 p_{i} =\frac{(1- f_{i}(\theta_{1},\theta_{2}, \ldots \theta_{m}))}{2^m}.
\end{eqnarray} 

First assume that after the $N$-qubit GHZ-basis measurement, we are able to create   $N$-party  entangled state and let us also focus on bipartite nonlocality. 
To consider the Bell-CHSH inequalities, $(N-2)$ parties perform local measurements having $2^{N-2}$ outcomes and the bipartite state, $\rho_{2}$ is  obtained.  For each measurement result, we find that two maximum eigenvalues of $T^{\dagger}T$ of $\rho_{12}$ are $p^4$ and $ \frac{p^{2N}\prod_{l=1}^{{N-2}}\sin^{2}(\theta_{l})}{(1-f_{i}(\theta_1,\ldots,\theta_{N-2}))^2}$. The average Bell-CHSH violation can be calculated as
\begin{eqnarray}
&\mathcal{L}_{NL}^{CHSH}\equiv\sum p_{i}BV(\rho_{i})\nonumber\\
& =  p^4 + \max_{\theta_{l},\phi_{l}}\sum_{i=1}^{2^{N-2}}\frac{p^{2N}\prod_{l=1}^{{N-2}}\sin^{2}(\theta_{l})}{2^{N-2}(1-f_{i}(\theta_1,\ldots,\theta_{N-2}))} -1.
\end{eqnarray}
For moderate number of parties i.e., for $N = 4, 5, 6$, we check that maximization in $\mathcal{L}_{NL}^{CHSH}$ is obtained when all the parties perform same measurements,  $\theta_i=\pi/2$ and it does not depend on $\phi_i$. We also notice that
\begin{equation}
 \max_{\{\theta_i\}}\sum_{i=1}^{2^{N-2}}\frac{\prod_{k=1}^{N-2}\sin(\theta_{k})}{2^{N-2}(1-f_{i}(\theta_1,\ldots,\theta_{N-2}))}|_{\theta_{i}=\pi/2} =1.
\end{equation}
If we assume that maximum is attained at $\theta_i=\pi/2$ for higher value of N as well,  we obtain the condition on the critical value of noise allowed to  the initial state so that the violation of the resulting state after steps 1 and 2 occurs. Specifically, we obtain $p_{cr}$ by solving the equation, given by
\begin{equation}
p^4  + p^{2N}-1 = 0.
\end{equation}
It implies that the initial Werner state should possess the mixing parameter, $p>p_{cr}$, which produces an output state violating Bell-CHSH inequality. For example, if $N=3$, we find that the critical value of the parent Werner state has to be greater than $0.869$ to obtain the violation of Bell inequality of the resulting state,. The value of $p_{cr}$ increases with the number of parties in the network. 

\subsection{Mermin-Belinskii-Klyshko nonlocality in star network: critical noise}
Let us now discuss the prescription by which  multipartite nonlocality among any prefix set of points in the star network can be  to established.
For N-qubit multipartite state, MBK operator can be written as \cite{dur'01}
  \begin{eqnarray}
  B_{N} = 2^{(N-1)/2} [e^{i\beta_{N}} |0\rangle\langle1|^{\otimes N}  +e^{-i\beta_{N}} |1\rangle\langle0|^{\otimes N}],
  \end{eqnarray}
 where  $ \beta_N =  \frac{\pi}{4N-4}$ and is obtained by putting $\sigma_{a_k}=\sigma_x$ and  $\sigma_{a^{\prime}_k}=\sigma_y$ for all values of k.
 Using this operator, we find the condition for violation of MBK inequality  by the N-party state in Eq. (\ref{sform}) as
  \begin{eqnarray}
 2^{(N-1)/2}P^{N}\cos{\beta_{N}}> 1 
  \end{eqnarray}
 which leads to
  \begin{eqnarray}
  p > p_{cr} = \frac{1}{2^{\frac{N-1}{2N}}(\cos{\beta_{N}})^{\frac{1}{N}}}.
  \end{eqnarray}
  
On the other hand,
in a collaborative network, $m$ parties perform local measurements,  and leave the network. Violation of MBK inequality can be calculated on remaining $N-m$-partite state, after performing optimization over local projective measurements by m-parties. The average violation of MBK inequality reduces to
 \begin{eqnarray}
\mathcal{L}_{NL}^{MBK}= \sum_{1}^{2^m} p_{i}|\mbox{Tr}[B_{N-m} \rho_{N-m}]|_{i} -1 \nonumber \\ =\sum_{1}^{2^m}\frac{2^{(N-m-1)/2}p^{N}[\cos({\beta_{N-m}-\sum_{1}^{m}\phi_{i}})\prod_{1}^{m}\sin\theta_{i}]}{2^m} -1 \nonumber\\ = 2^{(N-m-1)/2}p^{N}[\cos({\beta_{N-m}-\sum_{1}^{m}\phi_{i}})\prod_{1}^{m}\sin\theta_{i}] -1.\nonumber
\end{eqnarray} \\
From the analysis of small $N$, it can again be  shown to reach maximal value when \(\theta_{i} = \pi/2\) and  \( \beta_{N-m}=\sum_{1}^{m}\phi_{i}\).
Therefore, the violation of MBK inequality  of the resulting state leads to a maximum amount of noise permissible in the initial state. The condition reads as 
\begin{equation}
 2^{(N-m-1)/2}p^{N}> 1 
 \end{equation}
  implying
 \begin{equation}\label{pcrmk}
 p_{cr} = \frac{1}{(2^\frac{N-m-1}{2})^\frac{1}{N}}. 
  \end{equation}

\subsection{Noise threshold from functional Bell inequality: superadditivity}

Let us move  to a scenario where  violation of FB inequality of the output state of  $N-m$ parties in a star network  is investigated. As before, we are also interested to find out the  critical noise value of the initial Werner state leading to the violation of FB inequality in the multipartite state created after executing the protocol. 
We can choose the measurement operators in the $x-y$-plane of the Bloch sphere to calculate the violation of FB inequalities for states shared between $B_{i}$s i.e, 
\begin{eqnarray}
G_{n}(\eta_{n})=|+,\eta_{n}\rangle\langle+,\eta_{n}| -|-,\eta_{n}\rangle\langle-,\eta_{n}|,
\end{eqnarray}
where, $|\pm ,\eta_{n}\rangle = |0\rangle \pm e^{i\eta_{n}}|1\rangle$.
Since, the state has only two off-diagonal terms and diagonal terms,  quantum mechanical prediction in this case reads as \cite{aditi'05}
\begin{eqnarray}
 C_{QM}(\eta_{1},\ldots \eta_{N}) =Tr(G_{1} \ldots G_{N}\rho_{N})=p^{N} \cos(\sum_{i=1}^{N}\eta_{i}), \nonumber
 \end{eqnarray} 
 and
 \begin{eqnarray}
 \left \| C_{QM} \right \|^{2} = \int d\eta_{1} \ldots d\eta_{N}(C_{QM}(\eta_{1}....\eta_{n}))^{2}=\nonumber\\p^{2N} \int_{0}^{2\pi} d\eta_{1} \ldots d\eta_{N}(1+\cos(2\sum_{i=1}^{N}\eta_{i}))/2\nonumber\\=p^{2N} \frac{(2\pi)^{N}}{2}.    \nonumber
  \end{eqnarray}\\
 Similarly, the inner product of $C_{QM}$ and $C_{LHV}$ takes the form,
 \begin{eqnarray}
 &\langle C_{QM}|C_{LHV}\rangle = \int_{0}^{2\pi} d\eta_{1} \ldots d\eta_{N}C_{QM}C_{LHV} \nonumber\\&=\int_{0}^{2\pi} d\eta_{1} \ldots d\eta_{N}\int d\lambda \rho(\lambda)\prod_{n=1}^{N}\mathbb{I}_{n}(\eta_{n},\lambda)p^{N} \cos(\sum_{j=1}^{N}\eta_{j} )\nonumber \\& 
 \leq p^{N} 4^{N},
\end{eqnarray} 
where we have used the fact that \cite{zukowski'93}
\begin{equation}
\int_{0}^{2\pi} d\eta_{1}...d\eta_{N}\int d\lambda \rho(\lambda)\prod_{n=1}^{N}\mathbb{I}_{n}(\eta_{n},\lambda)\cos(\sum_{j=1}^{N}\eta_{j}) \leq 4^N. 
\end{equation}
When $\left \| C_{QM} \right \|^{2}$ is greater than $ p^{N} 4^{N}$, the state violates the FB inequality which leads to the threshold noise of the initial state, given by
  \begin{eqnarray}
  p^{N}  \geq 2\times (2/\pi)^{N}.\nonumber \\
   p_{cr}  = 2^{\frac{1}{N}}\times (2/\pi).
  \end{eqnarray}
It was shown in \cite{aditi'05} that the resulting multipartite state after the central GHZ measurement by Alice exhibit nonlocality even if the initial state does not -- the feature was called as superadditivity  which   is revealed when $N\geq7$.
\begin{figure}
\includegraphics[height=6.0cm,width=8.5cm]{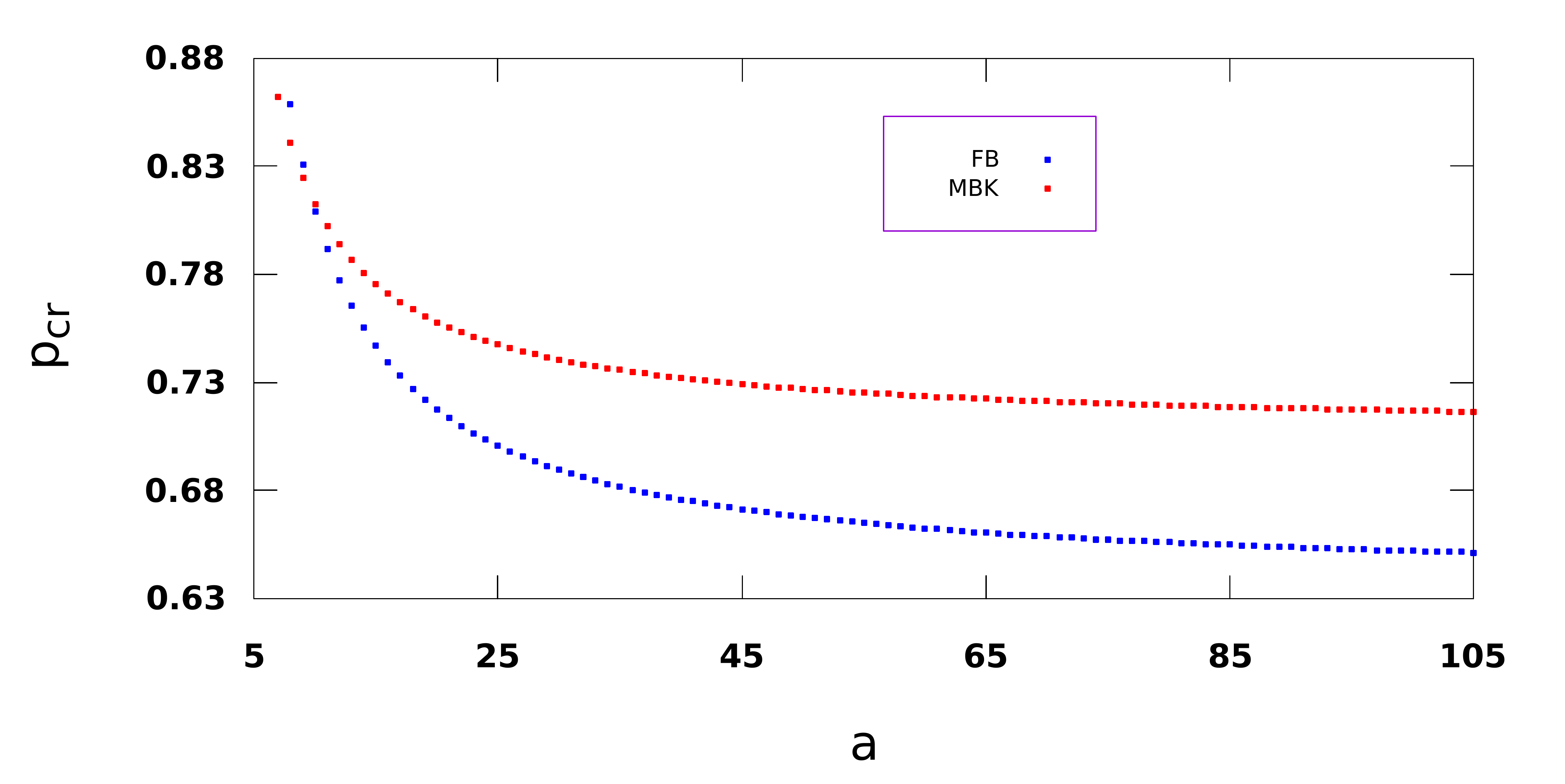} 
\caption{(Color online) Superadditivity in violation of local realism in star network with local measurements. Variation of the threshold noise allowed in the initial state, $p_{cr}$ (vertical axis) decreases with the number of copies of the initial state, $N$ (horizontal axis). Number of local measurements are fixed. Violation of MBK and FB inequalities are calculated. Clearly, we see that  to obtain violation via FB inequality of the final state, $\rho_{W}$ state can have $p<\frac{1}{\sqrt{2}}$.}
\label{fig:blockfunctional}
\end{figure}

We will now show that such a superadditivity of nonlocality can also be obtained in a collaborative star network. After local measurements by $m$ number of $B_{i}$s, the quantum mechanical correlation among $N-m$ parties post-selecting upon $\rho_{N-m}$ is given by
\begin{eqnarray}
  C_{QM}(\eta_{1}....\eta_{N-m})_{i} =Tr(G_{1}...G_{N-m}\rho_{N-m})\nonumber\\=\frac{p^{N}[\prod_{1}^{m}\sin\theta_{i}] \cos(\sum_{i=1}^{m}\phi_{i}-\sum_{i=1}^{N-m}\eta_{i} )}{(1- f_{i}(\theta_{1},\theta_{2},...\theta_{m}))},
  \end{eqnarray}
 and its norm can be found as 
  \begin{eqnarray}
&\left \| C_{QM} \right \|^{2}_{i} = \int d\eta_{1},...d\eta_{N-m}(C_{QM}(\eta_{1},....\eta_{N-m})_{i})^{2}\nonumber\\
=&\frac{p^{2N}[\prod_{1}^{m}\sin^{2}\theta_{i}] \int d\eta_{1}...d\eta_{n}(1+\cos(2\sum_{j=1}^{m}\phi_{j}-2\sum_{k=1}^{N-m}\eta_{k}))}{2(1- f_{i}(\theta_{1},\theta_{2},...\theta_{m}))^{2}}\nonumber\\
=&\frac{p^{2N}[\prod_{1}^{m}\sin^{2}\theta_{i}] (2\pi)^{N-m}}{2(1- f_{i}(\theta_{1},\theta_{2},...\theta_{m}))^{2}}.
\end{eqnarray}
The averaged QM predictions, where averaging is done over $2^m$ outcomes can then be written as
\begin{eqnarray}
  &\sum_{i=1}^{2^{m}}p_{i}\left \| C_{QM} \right \|^{2}_{i} \nonumber\\&= p^{2N}[\prod_{1}^{m}\sin^{2}\theta_{i}] \frac{(2\pi)^{N-m}}{2}\sum_{i=1}^{2^{m}}\frac{1}{2^{m}(1- f_{i}(\theta_{1},\theta_{2},...\theta_{m}))^{2}}.\nonumber
  \end{eqnarray}\\
  Similarly, the individual inner product of $C_{QM}$ and $C_{LHV}$ at the $i$-th measurement outcome 
  \begin{widetext}
  \begin{eqnarray}
  &\langle C_{QM}|C_{LHV}\rangle_{i}=\int d\eta_{1},...d\eta_{N-m}\int d\lambda \rho(\lambda)\prod_{n=1}^{N-m}\mathbb{I}_{n}(\eta_{n},\lambda)\frac{p^{N}[\prod_{1}^{m}\sin\theta_{j}] \cos(\sum_{k=1}^{m}\phi_{k}-\sum_{l=1}^{N-m}\eta_{l} )}{(1- f_{i}(\theta_{1},\theta_{2},...\theta_{m}))} \nonumber\\=
  &\frac{p^{N}[\prod_{1}^{m}\sin\theta_{j}]}{(1- f_{i}(\theta_{1},\theta_{2},...\theta_{m}))}\int d\eta_{1}...d\eta_{N-m}\int d\lambda \rho(\lambda)\prod_{n=1}^{N-m}\mathbb{I}_{n}(\eta_{n},\lambda)[\cos(\sum_{k=1}^{m}\phi_{k})\cos(\sum_{l=1}^{N-m}\eta_{l}) +\sin(\sum_{i=1}^{m}\phi_{i})\sin(\sum_{i=1}^{N-m}\eta_{i})]\nonumber\\
  &\leq \frac{p^{N}[\prod_{1}^{m}\sin\theta_{i}]}{(1- f_{i}(\theta_{1},\theta_{2},...\theta_{m}))} 4^{N-m}[\cos(\sum_{k=1}^{m}\phi_{k})+\sin(\sum_{l=1}^{m}\phi_{l})],
  \end{eqnarray}
  and its averged value can be written  as
  \begin{eqnarray}
  &\sum_{i=1}^{2^m}p_{i}\langle C_{QM}|C_{LHV}\rangle_{i} \leq p^{N}[\prod_{1}^{m}\sin\theta_{i}]4^{N-m}[\cos(\sum_{k=1}^{m}\phi_{k})+\sin(\sum_{l=1}^{m}\phi_{l})]\leq p^{N}[\prod_{1}^{m}\sin\theta_{i}]4^{N-m}\sqrt{2}.
\end{eqnarray}
\end{widetext}
Again for small N, we find that the measurement gives the optimal violation when all the $\theta_{i}$s take the value $\pi/2$ and \(\phi_{i}\)s do not play a role.
Therefore, from the violation of the localized FB inequality, $\mathcal{L}_{NL}^{FB}$, we obtain the critical value of  the initial noise parameter, satisfying the condition, given by
 \begin{eqnarray}
 p^{N} > 2\sqrt{2}\times(\frac{2}{\pi})^{N-m},
 \end{eqnarray}
which gives
\begin{eqnarray} \label{pcrfb}
 p_{cr} = 2^{\frac{3}{2N}}\times(\frac{2}{\pi})^{\frac{N-m}{N}}.
  \end{eqnarray}

\begin{figure*}
\includegraphics[height=5.0cm,width=16.0cm]{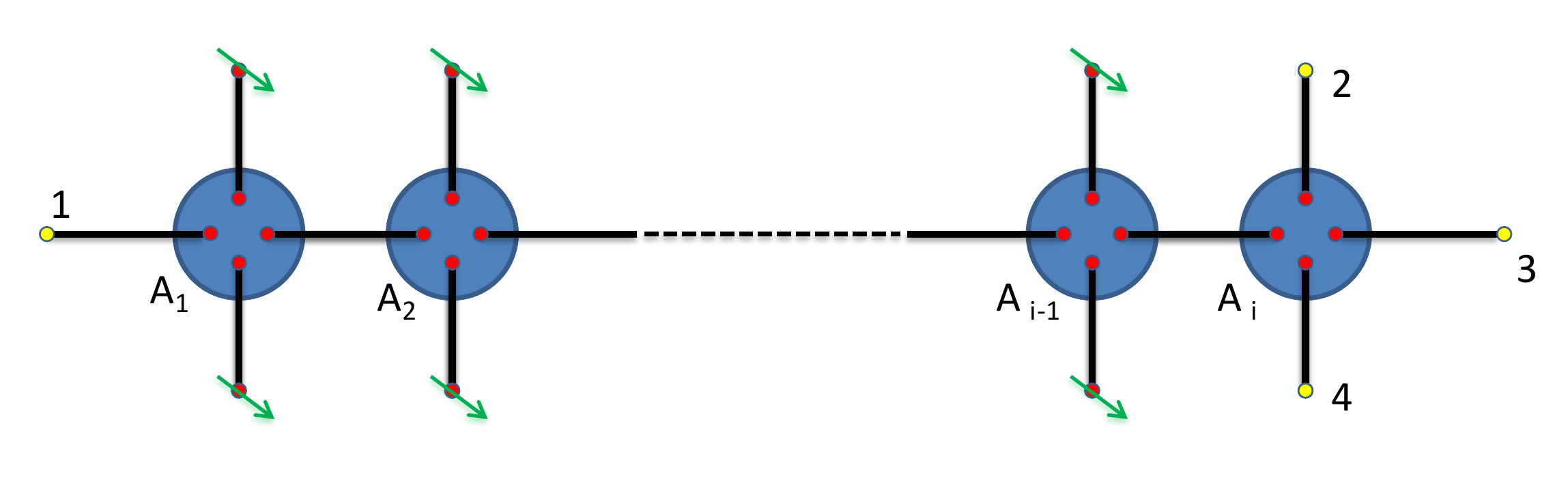} 
\caption{(Color online) Schematic diagram of an  one-dimensional lattice comprised of $\rho_W$ states with a fixed co-ordination number four. Four party  GHZ-basis and optimal local measurements are performed on the red qubits. We intend to produce a multipartite entangled state between the yellow qubits, marked as $1$, $2$, $3$ and $4$ whose entanglement can be verified by using the violation of local realism.}
\label{1dchain}
\end{figure*}

Comparing Eqs. (\ref{pcrmk}) and  (\ref{pcrfb}), we conclude that if we use FB inequality to detect multipartite nonlocality in the network, the  amount of noise allowed for obtaining  violation of the output state is higher than that of the MBK inequality, implying high robustness of localizable FB inequality aganist noise.  We also show that for a fixed N, $p_{cr}$ increases with the increase of number of measurements, $m$ while for fixed $m$, it decreases with $N$ as shown in Fig.  \ref{fig:blockfunctional}.  Moreover, we again report that the violation of $\mathcal{L}_{NL}^{FB}$ can be obtained even when  the initial  shared state can be non-Bell- violating, as depicted in Fig. \ref{fig:blockfunctional}, thereby also giving rise to superaditivity in violation of FB inequality  in a localized scenario (see \cite{aditi'05} for nonlocalized case).

\section{Distribution and detection of Bell nonlocality in Lattices}
\label{sec_network}
With the development of quantum communication protocols,  establishing and detecting nonclassical correlations in one-  and two-dimensional lattice network play an important role. We also investigate the minimal amount of entanglement required to obtain the quantum correlation among any prefix sites, detectable through the violation of Bell inequality after the entire protocol is completed. In one-dimensional network, we will also report that the output state obtained after the global and local measurement protocol can exhibit superadditivity in violation of Bell-type inequalities.

\subsection{One-dimensional Lattice}

Consider an one-dimensional lattice consisting of \(z\) number of nodes  with coordination  number \(a\) (the co-ordination number is defined as the number of connection in each node, for example, in Fig. \ref{1dchain}, a =4).
We call $A_{1}$, $A_{2}$, \(\ldots\) as nodes. The entire lattice is covered by bipartite quantum states, namely, Werner state.  First $A_{k}$s  perform joint measurements  and all the sites except those parties whom we want to connect, perform optimal local projective measurements. For example, in Fig. \ref{1dchain}, suppose we want to create an entangled state between $1$, $2$, $3$ and $4$, the local measurements are performed by all the sites except these.   In this situation, we are interested to find out whether the resulting multiparty state, shared between $1$, $2$, $3$ and $4$, violates Bell-type inequalities.  After the measurement by $A_{i}$s, we get a $(z(a-2)+2)$-party state, whose off diagonal terms only contribute in the violation of Bell-CHSH or MBK or FB inequalities and are given by 
\begin{eqnarray}
\langle 000\ldots \vert\rho_{z(a-2)+2}\vert111 \ldots \rangle = p^{z(a-1)+1}/2, \\
\langle111 \ldots\vert\rho_{z(a-2)+2}\vert000 \ldots \rangle = p^{z(a-1)+1}/2.  
\end{eqnarray}

\begin{figure}
\includegraphics[height=5.0cm,width=9.0cm]{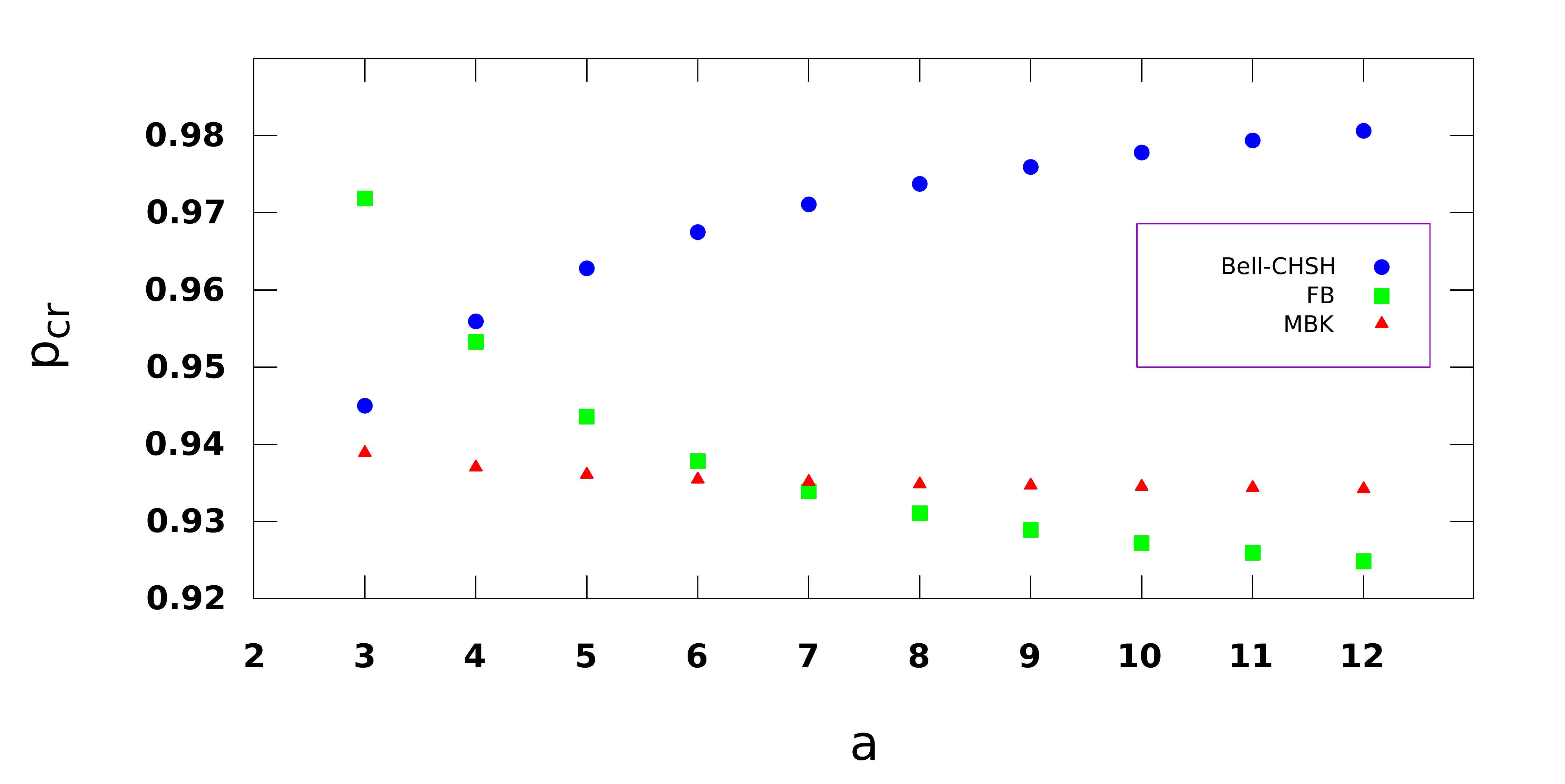} 
\caption{(Color online) Variation of the critical value of the noise parameter of the initial state  against the co-ordination number, $a$, of a chain. The abscissa and ordinate respectively represent $p_{cr}$ and $a$. We fix $z=5$. For fixed number of nodes, $p_{cr}$ increases with $a$ for obtaining  violation of Bell CHSH inequality (circle) while it decreases when  MBK (square),  FB (triangle) inequalities are considered.}
\label{pcrafunctional}
\end{figure}

\begin{figure}
\includegraphics[height=5.0cm,width=9.0cm]{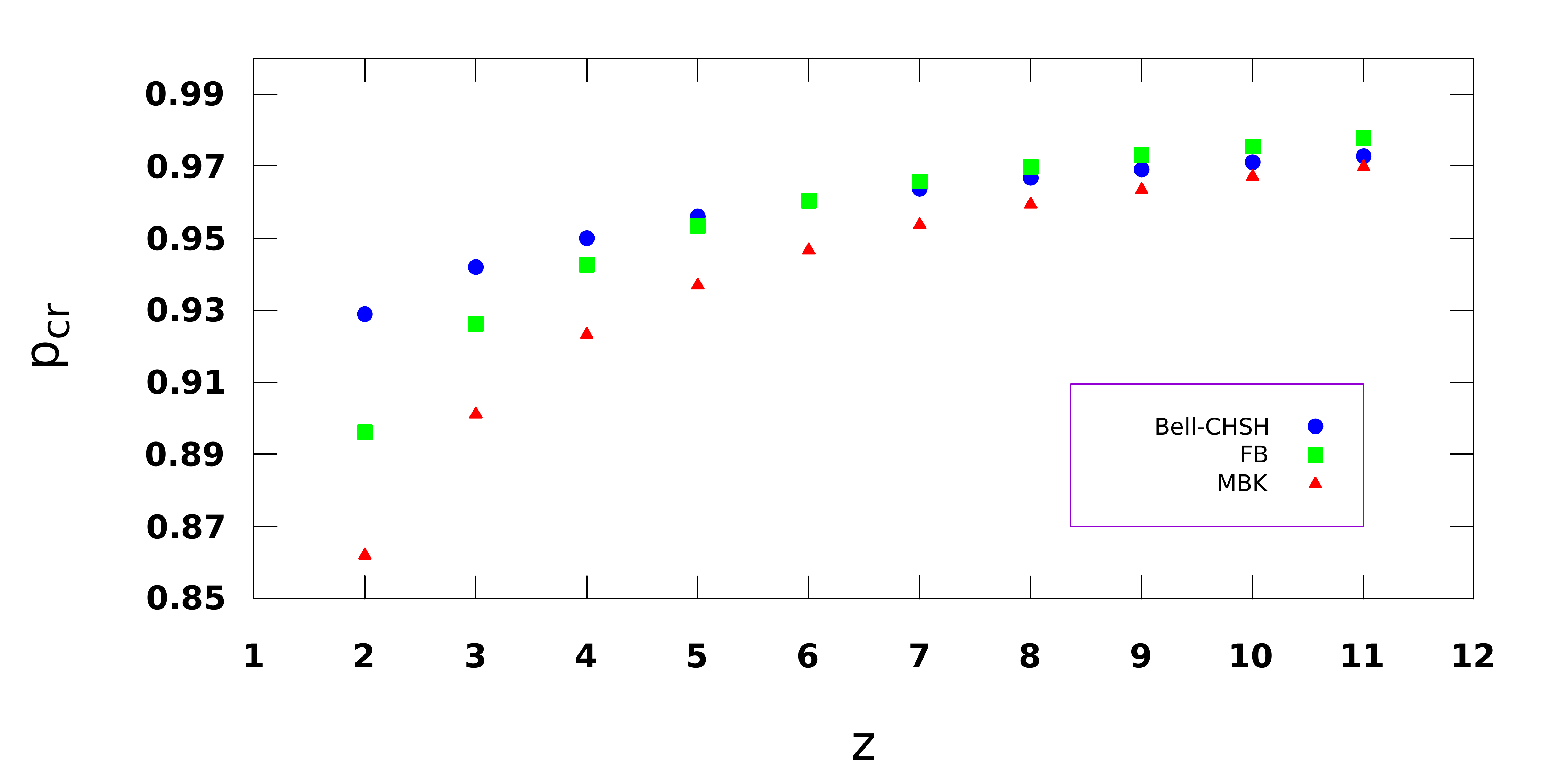} 
\caption{(Color online) $p_{cr}$ ($y$-axis) obtained by considering Bell-CHSH, MBK and FB inequalities with the increase of the number of nodes, $z$ ($x$-axis). Here we have  $a=4$. Other specifications are same as Fig. \ref{pcrafunctional}. }
\label{fig:pcrzfunctional}
\end{figure}

{\bf Spreading Bell nonlocality:} To obtain a two party state in two distant locations, say, between $1$ and $2$ in Fig. \ref{1dchain}, we perform measurements on remaining $z(a-2)$ parties having $2^{z(a-2)}$ outcomes. The average violation of Bell-CHSH inequalities can be calculated as 
\begin{eqnarray}
\sum p_{i}BV(\rho_{i}) =& p^6 + \sum_{i=1}^{2^{z(a-2)}}\frac{p^{2(z(a-1)+1)}\prod_{k=1}^{2^{z(a-2)}}\sin^{2}(\theta_{i})}{2^{z(a-2)}(1-f_{i}(\theta_1,...,\theta_{z(a-2)}))} \nonumber\\& -1.
\end{eqnarray}
As in the previous cases, average violation of Bell-CHSH inequalities attains its maximal value for $\theta_{i}= \pi/2$ and does not depend on $\phi_{i}$, which we check for a small lattice size. Therefore, the maximum amount of noise permissible for  the initial state can be obtained from 
\begin{eqnarray}\label{bell_1d}
p^6 + p^{2[z(a-1)+ 1]} -1 > 0.   
 \end{eqnarray}\\
From the above equation, it is clear that $p_{cr}$ depends both on $z$ as well as $a$. For a fixed $z$, we observe that $p_{cr}$ obtained from the violation of Bell-CHSH inequality increases with $a$ and the same trends persist when $a$ is fixed and $z$ is varying as depicted in Figs.  \ref{pcrafunctional} and \ref{fig:pcrzfunctional}.  

{\bf Spreading multipartite nonlocality:} Let us move to the violation of the averaged MBK inequality between the first  site connected to the first node  and $(a-1)$ parties of the last node of this chain. To establish such a connection,  the rest sites, i.e., $(z-1)(a-2)$ number of  parties perform optimal local measurements, which results a $\mathbf{a}$-party state.  The localizable MBK violation in a $\mathbf{a}$-party state reads as
\begin{eqnarray}
 \sum_{1}^{l} p_{i}|\mbox{Tr}[B^{a} \rho_{a}]|_{i} -1 \nonumber \\
 = \sum_{1}^{l}\frac{2^{(a-1)/2}p^{z(a-1)+1}[\cos({\beta_{a}-\sum_{1}^{l}\phi_{j}})\prod_{1}^{l}\sin\theta_{i}]}{l} -1 \nonumber\\
 =2^{(a-1)/2}p^{z(a-1)+1}[\cos({\beta_{a}-\sum_{1}^{l}\phi_{j}})\prod_{1}^{l}\sin\theta_{i}] -1,\nonumber
 \end{eqnarray}
 with
 \begin{eqnarray}
 l= 2^{(z-1)(a-2)},\nonumber
\end{eqnarray} \\
which can again be shown to be maximized when $\theta_{i} = \pi /2$ and $\beta_{a}=\sum_{1}^{l}\phi_{i}$.
 The state violates MBK inequality when p satisfying the  condition, given by
\begin{eqnarray}
 2^{(a-1)/2}p^{z(a-1)+1} > 1, 
 \end{eqnarray}
and therefore , we get 
\begin{eqnarray}
p_{cr} = 2^{\frac{1-a}{2z(a-1)+2}}.
\end{eqnarray}

{\bf Violation of functional Bell inequality and superadditivity:} Similar consideration also leads to  $p_{cr}$ using FB inequality between the first site of the first node  and $(a-1)$ parties of the last node. It reads as 
\begin{eqnarray}
 p^{z(a-1) +1} > 2\sqrt{2}\times(\frac{2}{\pi})^{a} \nonumber\\
 \Rightarrow p_{cr} = 2 ^{\frac{3}{2z(a-1) +2}}\times(\frac{2}{\pi})^{\frac{a}{z(a-1) +1}} .
  \end{eqnarray}
 Interestingly in the multipartite case, the threshold value of noise of the initial state decreases with $\mathbf{a}$ for fixed $z$. With the moderate value of the coordination number, $p_{cr}$ obtained from the violation of FB inequality decreases much faster than that of the MBK inequality (see Fig. \ref{pcrafunctional}). However for fixed $\mathbf{a}$, $p_{cr}$ increases with the increase of number of nodes $z$ (see Fig. \ref{fig:pcrzfunctional}).
Without local measurements on sites, the multiparty states violates FB inequality when the initial state has  
\begin{eqnarray}
 p > p_{cr} = 2 ^{\frac{1}{z(a-1) +1}}\times(\frac{2}{\pi})^{\frac{z(a-2)+2}{z(a-1) +1}}. 
\end{eqnarray} 
Interestingly,  superadditivity of nonlocality can be observed in this scenario with $a \geq 6$ for any arbitrary $z$. Specifically, if  an one-dimensional lattice having a fixed number of nodes is considered, we find that we require  coordination number to be \emph{six} to obtain an output state which can violate FB inequality   starting with non-violating Werner states. 

On the other hand, in a chain with $a=6$ and for a fixed  number of local measurements, we find minimum  number of nodes, $z$, required to exhibit the superadditivity in violation. Notice that with  the increase of $m$, $z$  increases to show superadditivity.  With $m$ number of local measurements, $p_{cr}$ is modified as
\begin{eqnarray}
 p_{cr} = 2 ^{\frac{3}{2z(a-1) +2}}\times(\frac{2}{\pi})^{\frac{z(a-2)+2-m}{z(a-1) +1}}. 
 \label{eq_pcr_FBsix}
\end{eqnarray} 
 For example, if we restrict ourselves to local measurements at ten sites, we find that $z\geq 69$ leads to superadditivity.

\subsection{Two-dimensional Lattice}

 \begin{figure}
\includegraphics[height=5.0cm,width=9.0cm]{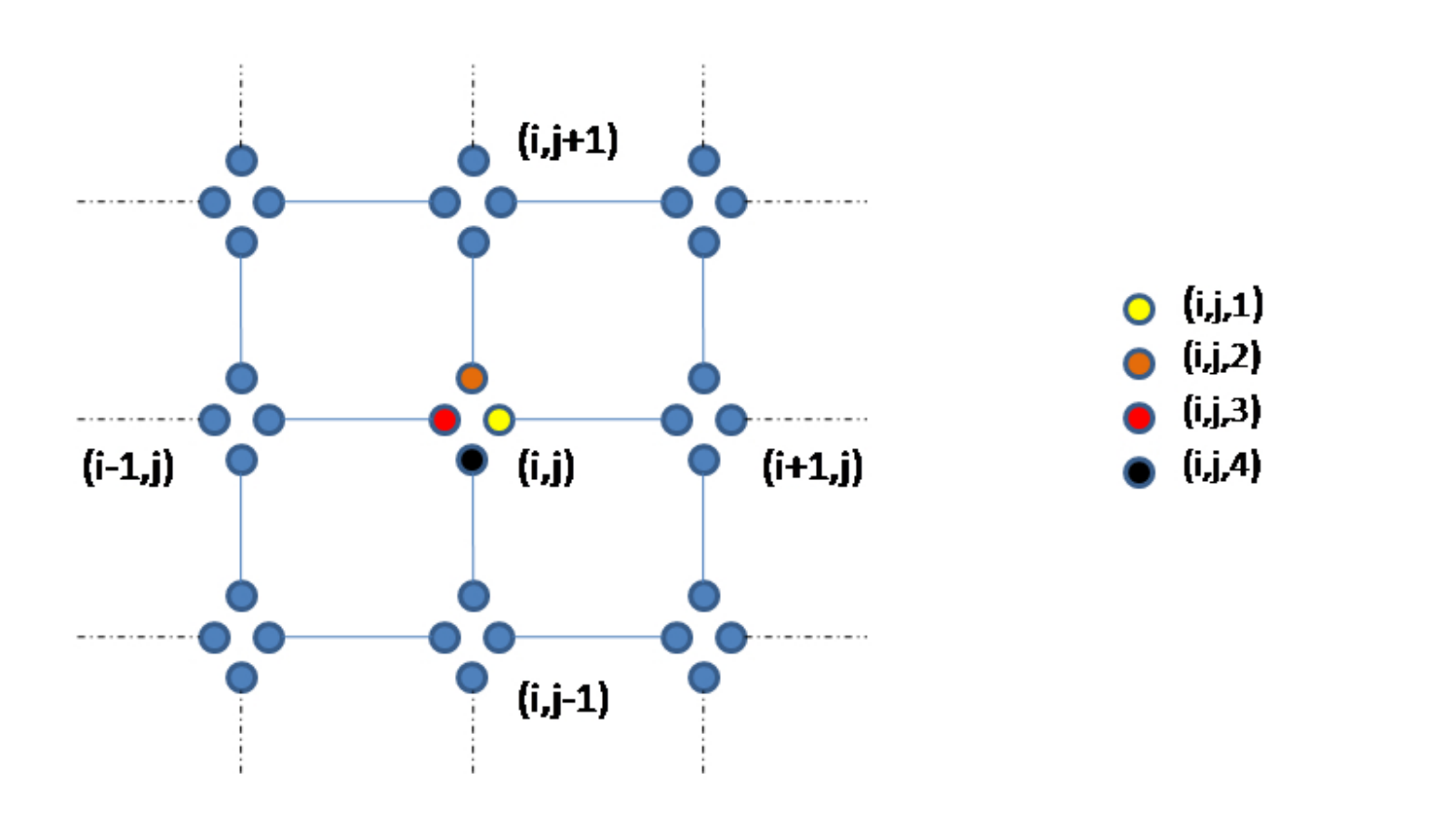} 
\caption{(Color online) Each node can be identified as $(i,j)$. While we require another co-ordinate to indicate sites in each node. Each site can be spotted by $(i,j,q)$ .Nearest nodes connected to $(i,j)$ is shown.}
\label{numbering2d}
\end{figure}

\begin{figure}
\includegraphics[height=6.0cm,width=8.0cm]{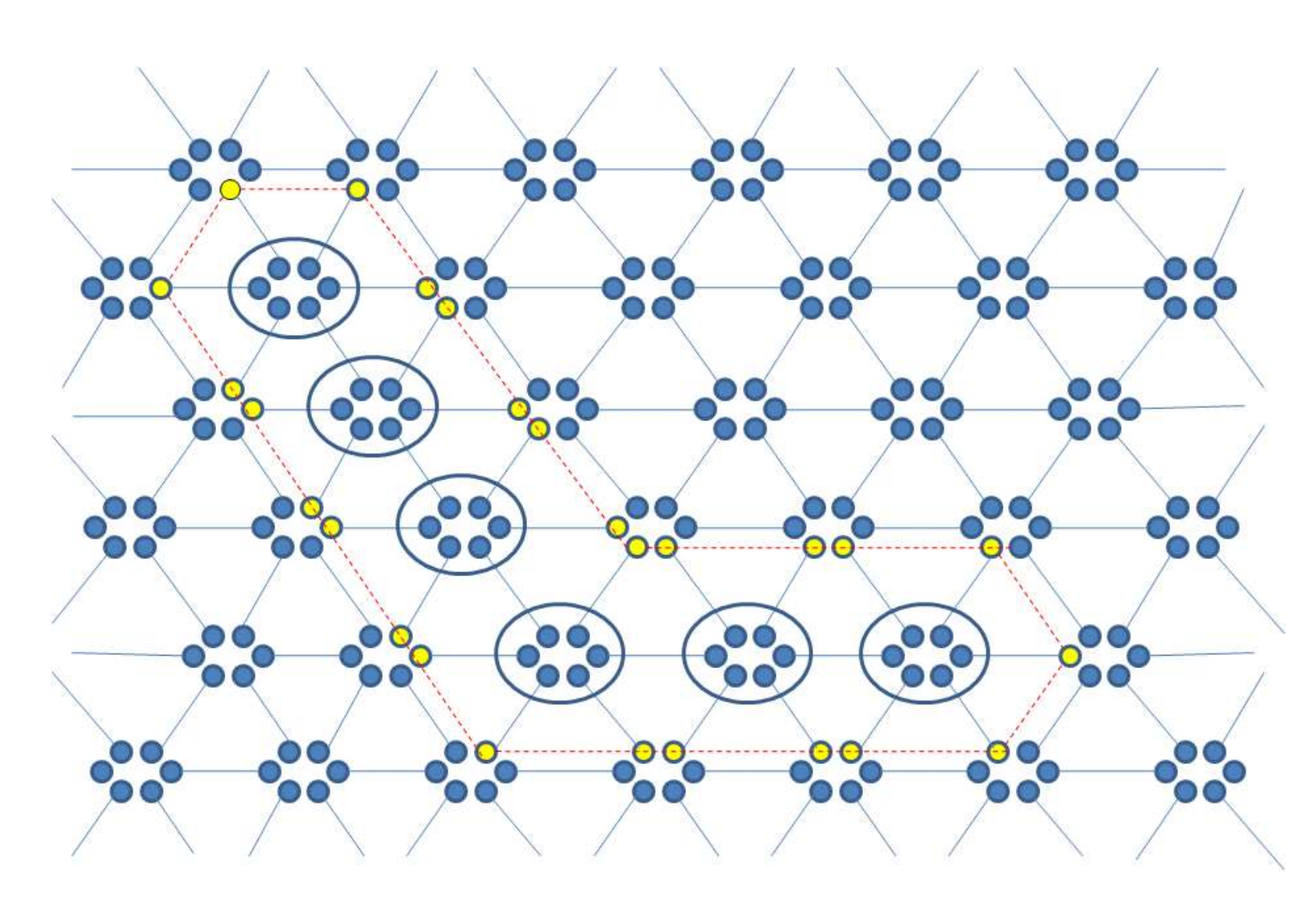} 
\caption{(Color online) A triangular lattice with coordination number six. Creation of multipartite state after joint measurements are shown with dotted line.}
\label{2d_triangle}
\end{figure}

We consider two kinds of lattices having two different coordination numbers, namely square (with $a=4$) and triangular (having $a=6$) lattices (see Figs. \ref{numbering2d} -  \ref{2d_triangle}). In both the cases, we prescribe an algorithm to share entangled state between any two or more distant points of the network.
\subsubsection{Square Lattice}
\begin{figure}
\includegraphics[height=6.0cm,width=8.0cm]{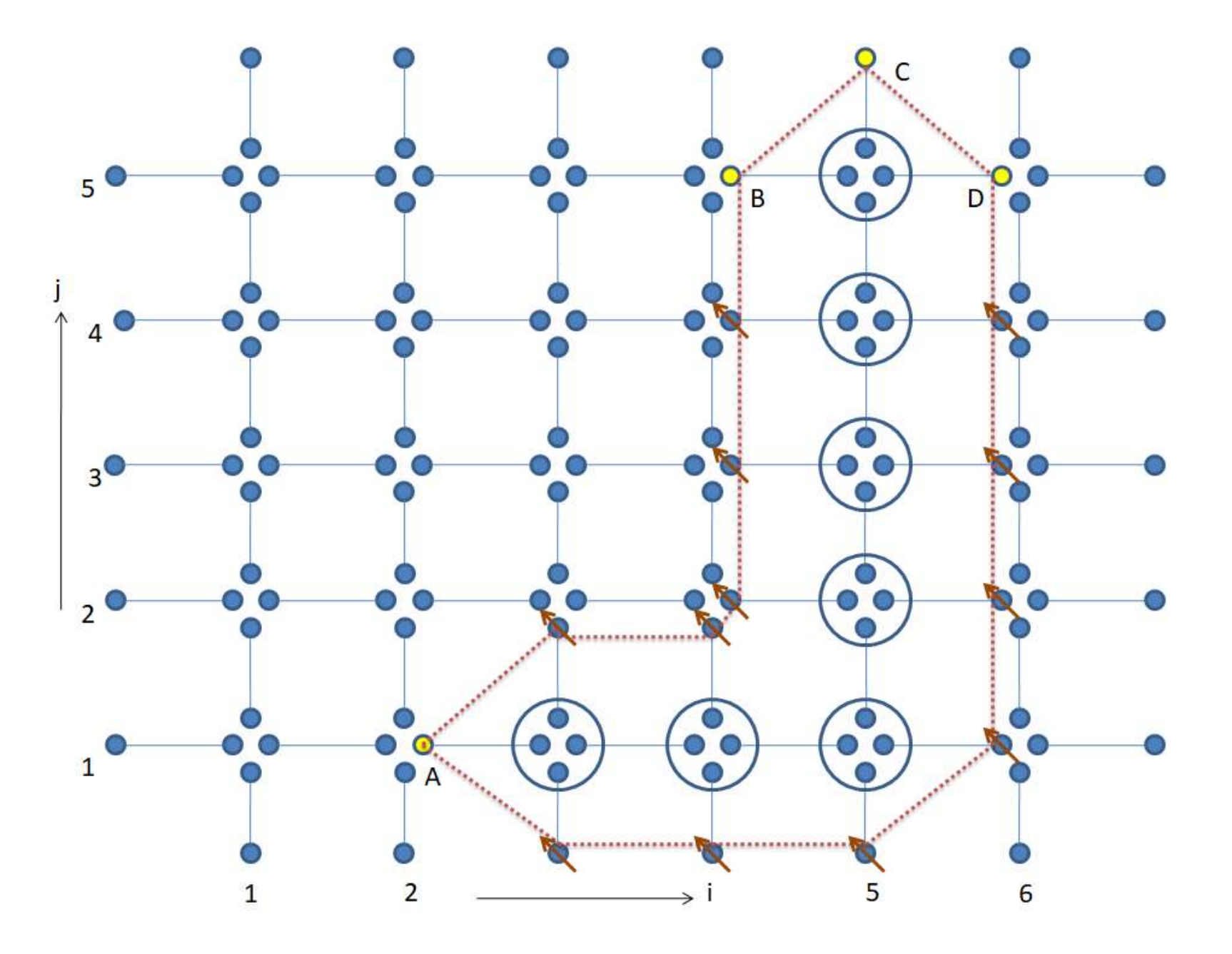} 
\caption{(Color online) A square lattice. To establish entanglement between A , B, C and D, the joint and local measurements are marked by circle and arrow respectively.   }
\label{2d_square}
\end{figure}

Suppose in a square lattice, as in Fig. \ref{2d_square}, , our aim is to have a bipartite state between  A and D. The prescription for establishing the connection is as follows: 
 \begin{enumerate}
\item 
Let us first fix the notation used to describe nodes. Firstly, the network is in a 2D plane and hence position of any node can be described by using two numbers. Since $a=4$, each node consists of four parties which are eventually connected to four different nodes, in four directions. After specifying the position of a node, another number is required to fix the position of the party within this node as shown in Fig.   \ref{numbering2d}. For an example, the number $(i, j, q)$  denotes the    q-th party in the $(i, j)$-th node, where q can be $1$, $2$, $3$ and $4$.
\item Suppose we choose two sites of different nodes, given by $({i},{j},{q})$,$({i}',{j}',{q}')$. To connect these two parties, we can apply the following  rule:\begin{enumerate}
\item First, we have to find the nearest node connected to $({i},{j},{q})$ and $({i}',{j}',{q}')$. Depending on the value of ${q}$,  we can specify the nearest connected node to  $({i},{j},{q})$ as follows :
\begin{eqnarray}
q=1: ({i+1},{j})  \nonumber\\
q=2:   ({i},{j+1}) \nonumber\\
q=3:     ({i-1},{j})  \nonumber\\
q=4:        ({i},{j-1}) \nonumber\\
 \end{eqnarray}\\
Similarly, nearest nodes to any point can be identified (see Fig. \ref{numbering2d}).

\item Suppose ${q} =1$ and ${q}'= 3$. The nearset connected nodes are then $({i}+1,{j})$ and $({i}'-1,{j}')$. Let us perform four party GHZ-basis measurements (marked as circle in Fig. \ref{2d_square}) from the node denoted by $(i+1, j)$  to $(i^{\prime}-1, j)$, denoted by $(i+1, j) \rightarrow (i^{\prime}-1, j)$. Then measurements are performed in a direction given by $(i^{\prime}-1, j) \rightarrow (i^{\prime}-1, j^{\prime})$. After these measurements, we get a chain which establishes multipartite entangled state between $(i, j, q)$ and ($i^{\prime}, j^{\prime}, q^{\prime})$ and hence we can apply the results obtained in previous subsection. To get the violation of local realism between A and D, one has to measure locally in an optimal basis at all the sites except A and D. Similarly we can also create multiparty entangled state between them by performing GHZ-basis measurements  in the direction 
$(i+1, j) \rightarrow (i+1, j^{\prime}) \rightarrow (i^{\prime}-1, j^{\prime})$.
To obtain the violation of Bell-CHSH inequality between A and D, the maximal noise allowed in the Werner state can be obtained by using Eq. (\ref{bell_1d}). Specifically, after replacing z and a as
\begin{eqnarray}
 z = (i_{2} - i_{1}) + (j_{2} - j_{1} +1); \,\, a =4 
 \end{eqnarray}
 where $(i_{1},j_{1})$ and  $(i_{2},j_{2})$ are nearest connected node of $({i},{j},{q})$, $({i}',{j}',{q}')$ respectively.
\end{enumerate}
\end{enumerate}


 \begin{center}
 	\begin{table}
 		\begin{tabular}{|*{7}{c|}}
\hline
$(i,j,q)$ & $ (i \prime,j \prime, q \prime) $& $(i_{1},j_{1})$ &  $(i_{2},j_{2})$ & $z$ & $a$ & $p_{cr}$\\
\hline
$(2,1,1)$ & $ (6,5,3) $& $(2,2)$ &  $(5,5)$ & $7$ & $4$ & $0.9658$\\
\hline
$(2,1,1)$ & $ (3,3,2) $& $(2,2)$ &  $(3,4)$ & $4$ & $4$ & $0.9427$\\
\hline
$(2,1,1)$ & $ (3,2,4) $& $(2,2)$ &  $(3,2)$ & $2$ & $4$ & $0.8963$\\
\hline
\end{tabular}  
\caption{Example of $p_{cr}$ for spreading non-locality from a fixed point of a 2D network to three different points. }
\label{tab}
\end{table}
\end{center} 
The above prescription can also be used to generate multipartite entangled states when the entire lattice comprises of  $\rho_{W}$. Critical noise of the initial state leading to a multipartite state between A ,B, C and D which violates FB inequality is listed in Table \ref{tab}.

\normalsize 

\subsubsection{Triangular Lattice }
In the triangular  lattice, the coordination number is six. This geometry is considered since we show in 1D lattice having coordination number six is special. Hence such a lattice has potential to show superadditivity in violation of local realism. Suppose, we want to create a multipartite state between sites, marked with yellow dots in  Fig. \ref{2d_triangle}. For generating such a multipartite entangled state, six joint measurements have to be performed in six-qubit GHZ-basis. From the results obtained for 1D lattice, it is clear that even if the initial Werner state has $p<\frac{1}{\sqrt{2}}$, the final state still violates FB inequality, thereby showing superadditivity in violation of local realism.  By using Eq. (\ref{eq_pcr_FBsix}), we find that the superadditivity can  be observed for a multiparty state having \emph{fifity three} sites, if we  perform local measurement on a single site of the lattice.  In particular, for a fixed number of local measurements, $m$, we can  find a minimum number of nodes required to exhibit superadditivity.

\section{conclusion}
\label{sec_conclu}
Establishing connection between two or more parties via producing entanglement between them  is essential to implement quantum information protocols in network. Generation of entanglement among arbitrary prefix set of points has to be guaranteed by using certain detection procedures. In this paper, we employed bipartite as well as multipartite Bell tests to certify entanglement device-independent way  in network of one- and two-dimensional square as well as triangular lattices which are initially covered by arbitrary number of noisy entangled states. In a two party scenario, we considered CHSH inequality while for  multipartite states, we evaluated MBK, a two-setting Bell inequality, as well as functional Bell inequality having continuous settings. We proposed  joint and local measurement-based method to establish entanglement in such a network with arbitrary number of nodes. For a fixed number of nodes and a fixed  coordination number, we found the entanglement content of the initial state required so that the resulting state violates a certain type of Bell inequality. Our method shows that the number of nodes, coordination number as well as joint and local measurements have a interplay in obtaining violation of Bell inequality.

We reported that in case of functional Bell inequality,  a method presented here can produce a state that violates Bell inequality although the initial state is not Bell-violating. In particular, we found that the minimum  coordination number required to activate such a superadditivity phenomena is six  in an one-dimensional lattice with an arbitrary number of nodes. Based on this result, we designed a protocol on a triangular lattice in which there  exists a final output state  violating functional Bell inequality after joint and local measurements although the initial states covering the lattice do not violate any local realism. Such a phenomena is absent in square lattice.  Our proposed architecture of connecting any prefix sites in a lattice can be a step  towards building the quantum internet.

\end{document}